\documentclass[12pt]{article}
\usepackage{amsmath}
\usepackage{amsfonts}
\usepackage{amssymb}
\usepackage{float}
\usepackage{soul}

\usepackage{subcaption}
\usepackage{apacite}
\usepackage{graphicx}
\usepackage{microtype}
\usepackage{sectsty}\allsectionsfont{\raggedright}

\newtheorem{theorem}{Theorem}

\newtheorem{corollary}[theorem]{Corollary}

\newtheorem{proposition}[theorem]{Proposition}

\begin{document}
\renewcommand{\refname}{REFERENCES}

\title{SURFACE GRAVITY OF ROTATING DUMBBELL SHAPES}
\author{Wai Ting Lam\thanks{Research partially supported by NSF grant DMS-0635607 and DMS-1814543.}, Marian Gidea\footnotemark[1], and Fredy R Zypman\thanks{Corresponding author: zypman@yu.edu. Partially supported by NSF grant CHE-1508085}
\\Yeshiva University, 2495 Amsterdam Avenue,\\ New York City, NY 10033}
\date{}
\maketitle

\begin{abstract}
We investigate the problem of determining the shape of a rotating celestial object -- e.g., a comet or an asteroid -- under its own gravitational field. More specifically, we consider an object symmetric with respect to one axis -- such as a dumbbell -- that rotates around a second axis perpendicular to the symmetry axis.  We assume that the object can be modeled as an incompressible fluid of constant mass density, which is regarded as a first approximation of an aggregate of particles.

In the literature, the gravitational field of a body is often described as a multipolar expansion involving spherical coordinates \cite{kaula1966theory}. In this work we describe the shape in terms of cylindrical coordinates, which are most naturally adapted to the symmetry of the body, and we express the gravitational potential generated by the rotating body as a simple formula in terms of elliptic integrals.
An equilibrium shape occurs when the gravitational potential energy and the rotational kinetic energy  at the surface of the body  balance each other out.
Such an equilibrium shape can be derived  as a solution of  an optimization problem, which can be found via the variational method. We give an example where we apply this method to a two-parameter family of dumbbell shapes, and find approximate numerical solutions to the corresponding optimization problem.

\end{abstract}

%\newpage

\section{INTRODUCTION}

Modeling the gravitational fields produced by celestial bodies in the Solar System has been of interest since the time of Isaac Newton, Alexis Claude Clairaut, and George Gabriel Stokes. Their work was focused on determining the shape of the Earth. The more general problem is to determine all possible equilibrium shapes of rotating, homogeneous fluid bodies.
Such equilibrium shapes are given by the condition that the total energy, i.e., the sum of the gravitational potential energy and the rotational kinetic energy, should have the same value at any point of unit mass on the surface of the body.
The rotational speed is a parameter of the problem, with different rotational speeds yielding different equilibrium shapes.
Maclaurin \cite{maclaurin1742treatise}  discovered a family of equilibrium shapes -- referred to as Maclaurin spheroids --, which are  ellipsoids that are symmetric with respect to the axis of rotation. As the rotational speed of the   body is increased, the family of spheroid equilibrium shapes branches off at some  value of the rotational speed,  giving rise to a family of   tri-axial ellipsoids, referred to as the Jacobi ellipsoids. Further, the family of Jacobi ellipsoids branches off into two families of equilibrium shapes at two distinct  values of the rotational speed.
One of the branches  consists of pear-shaped equilibrium figures, referred to as Poincar\'e figures.
The second branch consists of  dumbbell-shaped equilibria. The corresponding  branching  point  for this latter family was found in \cite{chandrasekhar1967post}. The dumbbell sequence was first computed in \cite{eriguchi1982dumb}. One should note here that these dumbbell shapes are not given by closed form equations, but they are computed numerically, for example via iterative methods.

Additionally, the gravitational fields of oblate planets have been of interest to understand the motion of satellites, in particular artificial satellites orbiting around the Earth \cite{vinti1966invariant,lara2020solution}.  More recently, general shapes have been considered with the aim of providing accurate dynamics of interacting gravitational bodies that can be monitored experimentally to high precision \cite{dirkx2019propagation,celletti2020resonances}.
In such problems, multipolar expansions  of the gravitational potential play a key role in the analysis.

The interest in modeling and computing the gravitational field created by shapes beyond spheroids is stimulated by the fact that many asteroids and comets have irregular shapes. Dumbbell shapes are among the shapes that have been observed for comets and asteroids, making them both astronomically and mathematically interesting.

Examples of astronomical dumbbells include the Jupiter  Trojan Asteroid 624 Hektor \cite{descamps2015dumb}, the comet 8P/Tuttle \cite{groussin2019spitzer}, the comet 103P/Hartley \cite{harmon2011radar}, and the transneptunian object 486958 Arrokoth/Ultima Thule \cite{amarante2020surface}. Astronomically, these dumbbell shapes can originate from fusion of ellipsoidal precursors, or from elongation.  We do not quest here into the mechanisms of formation, rather concentrating on the self-gravity of  bodies, and searching for possible shapes that can occur in practice.

%minimize the gravitational surface energy.

A related problem, of astrodynamics interest, is to understand the dynamics of an infinitesimal mass
(e.g, a spacecraft) near a dumbbell shaped asteroid (e.g., the Trojan Asteroid 624 Hektor), under the additional influence of other planetary bodies (e.g., Sun, Jupiter). See, e.g.,  \cite{burgos2020hill}.

For spheroid shapes the gravitational field can be expressed as a spherical harmonic expansion. However, this method is not particularly suitable for non-spheroidal bodies. Other types of expansions have been proposed. For instance, ellipsoid harmonic expansions \cite{romain2001ellipsoidal} are better fitted for bodies that can be approximated by an ellipsoid rather than by a sphere.

In this work we provide a general expression for the gravitational potential produced by a body of constant mass density, whose boundary surface is described by revolving the graph of a single-valued function about an axis, which becomes the symmetry axis of the body.

Using cylindrical coordinates, which are adapted most naturally to the symmetry of the problem, we obtain a general formula for the gravitational potential as a one-dimensional integral of a closed form expression given in terms of elliptical functions.

When the rotation of the body about an axis perpendicular to the symmetry axis is considered,  the total energy of a unit mass particle at the surface of the body is expressed as the sum of the  gravitational potential energy and  the rotational kinetic energy.

Then we formulate the problem of finding shapes of axisymmetric rotating bodies that are approximate  equilibrium shapes of the total energy, in the sense that the surface of the body is as close as possible to a  level set of the total energy function. We formulate this problem as a  two-dimensional optimization problem, which can be solved numerically using variational methods. We illustrate this approach via an example consisting of a two-parameter family of dumbbell shapes, for which we find numerical solutions to the corresponding optimization problem.

The physical justification that the surface of an equilibrium shape  is approximately  a  level set of the total energy function  relies on viewing the body as either a solidifying fluid, or a collection of solid particles. In the case of asteroids, this assumption is partially justified by the fact that many small bodies in the solar system are believed to be rubble piles, that is, collections of smaller particles. There are models that analyze in detail  the  granular structure of asteroids, and study the tidal stress corresponding to different particle shapes; see, e.g.,  \cite{goldreich2009tidal}. However, numerical simulations show that such granular structures  preferentially assume shapes that are close to fluid equilibrium shapes \cite{tanga2009thermal}. Nevertheless,  perfect equilibrium fluid shapes are not attained since the bodies are not truly  fluid but subject to some level of inter-particle friction. The fluidity hypothesis-based approach was recently used in \cite{descamps2015dumb} to find, through an iterative scheme, a family of shapes that fit the observed light curves\footnote{ Light curves are measurements of the brightness of a celestial body as a function of time.  They  are used for example to determine the shape, rotation period, as well as other parameters of an asteroid \cite{kaasalainen2001optimization}.} of some small bodies in our Solar System, for example of the Trojan Asteroid 624 Hektor.

The paper is divided into three main sections.  Section~\ref{section1}  provides a  formula for the gravitational potential of a   solid of revolution in terms of elliptical functions. This formula provides an economical method to exactly compute the gravitational field inside and outside the body.
Section \ref{section2} gives an integral equation that any rotating such body, assuming an equilibrium shape, must satisfy.  %The strength of this result lies in its simplicity.
This result serves as a practical starting point to explore possible equilibrium shapes that can be attained from suitably chosen  families of profiles. Section \ref{section3} presents an application to a parameterized family of dumbbells, which uses an optimization process to identify the best fit shapes.  Section~\ref{section4} presents the conclusions.

\section{AXISYMMETRIC BODY IN ITS OWN GRAVITATIONAL FIELD}
\label{section1}
\begin{figure}[h]
\begin{center}
{\includegraphics[width=0.45\linewidth]{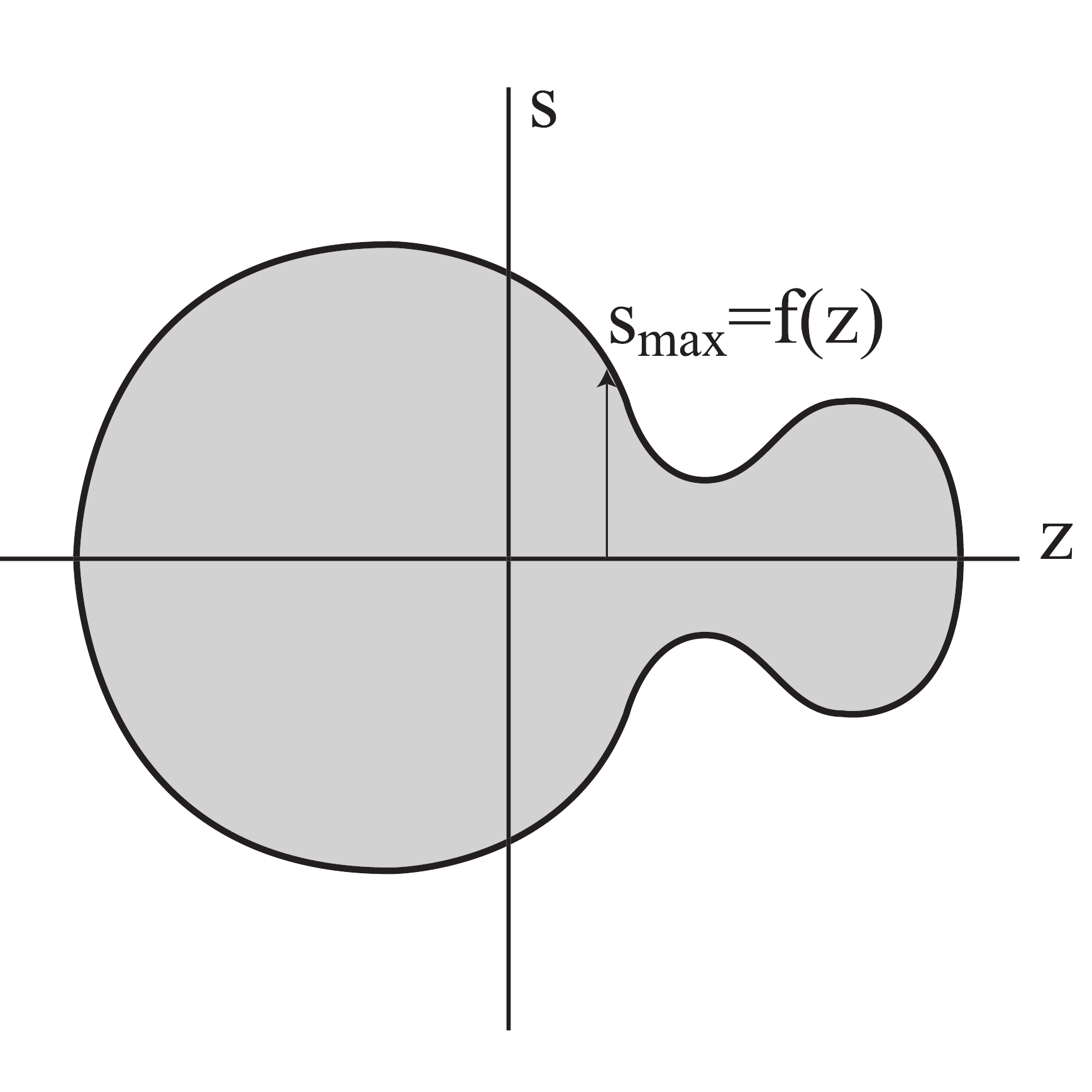}}
{\caption{Sagittal section of body of revolution around the $z$-axis.  All points inside the object are within
$-z_0\leq z\leq z_1$.  For a fixed $z$, the points are on a circular disc of radius $s_{\rm max}$.}\label{fig:Sagittalsect}}
\end{center}
\end{figure}

We consider an axisymmetric body relative to the $z$-axis, obtained by revolving the graph of single-valued function $s_{\rm max}=f(z)\geq 0$ about the $z$-axis. The origin of the $(z,s)$-coordinate system is set at the center of mass.   See Figure~\ref{fig:Sagittalsect}.

In this section,  we consider shapes that have only this symmetry.  In Section \ref{section3}, we will restrict to a family of shapes that have an additional symmetry, namely that  the graph of the function $f(z)$ is symmetric about the $s$-axis.

%The figure shows a shape that is also symmetric with respect to the $s-$axis.
 %While our main result, equation \eqref{prop:1},does not require this additional symmetry, we focus mainly on that case since, upon rotation around the $s-$axis, the minimum shape will be attained within the higher symmetry sub-family.
 In cylindrical coordinates $(s,\phi,z)$, the inside of the body is defined by $ -z_0\leq z\leq z_1$, $0 \leq s \leq s_{\rm max}=f(z)$, $0 \leq \phi \leq 2\pi$.  The function $f(z)$ defines the shape of the body.
	The points $\vec{r'}$ inside the body produce a gravitational potential at a point $\vec{r}=z\hat{z}+f(z)\hat{s}$ on the surface that is given by
\begin{equation}\label{UG}
U_{\rm G}(\vec{r})=-G\int_{\rm Body} \frac{\rho\, d^3 \vec{r'}}{\lvert \vec{r}-\vec{r'}\rvert},
\end{equation}
where $\rho$ is the  constant mass density of the object, $G$ is the universal gravitational constant, and the caret stands for the unit vectors in the axial and transverse directions.
We note that the integrand in \eqref{UG} is undefined when $\vec{r'}=\vec{r}$, yet the value of the potential at any point $\vec{r}$ on the surface of the body is well defined owing to the three-dimensional nature of the body.

	%If the object rotates, for example around the $s$-axis, then an additional rotational potential $U_R$, which is proportional to the square of the angular velocity, is added the gravitational potential. See Section~\ref{section2}.

	We begin by considering the explicit form of equation \eqref{UG} in cylindrical coordinates \cite{skelton1982free}, \cite{conway2000analytical}:
\begin{equation}\label{UGexplicit}
U_{\rm G}=-G \rho \iiint dz' ds' d\phi' \,s'\sum_{m=-\infty}^{+\infty} \int_{0}^{+\infty} dk\, e^{im(\phi'-\phi)} J_m(ks)J_m(ks')e^{-k\lvert z-z' \rvert}.
\end{equation}
%where $\phi$, $s$ and $z$ represent cylindrical coordinates.
Here we will use the primed variables to refer to the sources and the unprimed for the observation point. The integrals in equation \eqref{UGexplicit} are in the variables $z'\in[-z_0,z_1]$, $s'\in[0,f(z')]$, and $\phi'\in[0,2\pi]$, respectively.

The triple integral in \eqref{UGexplicit} can be rearranged as a sum of an infinite series as follows
\begin{equation}\label{UGrearrage}
\begin{split}
\sum_{m=-\infty}^{+\infty} \int_{0}^{+\infty} dk\, \int_{-z_0}^{z_1} dz'\, e^{-k|z-z'|}\int_{0}^{f(z')}ds'\, s' J_m (ks) J_m(ks')\int_{0}^{2\pi} d\phi'\,e^{im(\phi'-\phi)},
\end{split}
\end{equation}
where the variable $s$ is evaluated at the surface of the body.
%The integral in $z'$ is between $-z_0$ and $z_1$.
The integral in $\phi'$ in equation \eqref{UGrearrage} is equal to $2\pi \delta_{m0}$, where $\delta$  is the Kronecker function.  Thus, equation \eqref{UGexplicit} becomes
\begin{equation}
U_{\rm G} = -2 \pi G \rho \int_{0}^{+\infty} dk \int_{-z_0}^{z_1} dz' e^{-k\lvert z-z'\rvert} \int_0^{f(z')} ds' \,  s' J_0(ks) J_0(ks') .
\end{equation}

We now recall the identity \cite{watson1922treatise}
\begin{equation}
s' J_0(ks') = \frac{1}{k}\frac{d}{ds'}[s' J_1(ks')]
\end{equation}
which allows us to evaluate the integral in  $s'$ and obtain
\begin{equation}\label{UGevaluates'}
U_{\rm G} = -2 \pi G \rho \int_{0}^{+\infty} dk \int_{-z_0}^{z_1} dz' e^{-k\lvert z-z'\rvert} \frac{J_0(ks)}{k} f(z') J_1(kf(z'))
\end{equation}
where we have used that $J_1 (0)=0$.

For clarity, equation \eqref{UGevaluates'} is rearranged as
\begin{equation}\label{UGevaluates'2}
U_{\rm G} = -2 \pi G \rho \int_{-z_0}^{z_1} dz'\, f(z') \int_{0}^{+\infty} dk\, \frac{J_1(kf(z')) J_0(kf(z))}{k}e^{-k| z-z'|},
\end{equation}
where we have explicitly set $s=f(z)$ to obtain the potential at the surface.

We now consider the integral
\begin{equation}\label{Yint}
\textbf{Y}(f(z'), f(z), |z-z'|)=\int_{0}^{+\infty} dk\, \frac{J_1(kf(z'))J_0(kf(z))}{k} e^{-k\lvert z'-z\rvert}.
\end{equation}

This integral can be viewed  as  the Laplace transform of a combination of Bessel and power functions, of the kind studied in  \cite{kausel2012laplace}.
Specifically, using the notation from \cite{kausel2012laplace},
\begin{equation*}\textbf{I}_{\alpha\beta}^{\lambda}(a,b,s):=\int_0^{+\infty} dx\, x^{\lambda}J_\alpha(ax)J_\beta(bx)e^{-sx},\end{equation*}
for
\begin{equation}\label{eqn:letters}\begin{split}
\alpha=1,\, \beta=0,\, \lambda=-1,\, x=k,\, a=f(z'),\, b=f(z),\textrm{  and }s=|z'-z|
\end{split}\end{equation}
we have that
the right-hand side of equation \eqref{Yint} equals
\begin{equation*}\textbf{I}_{\rm 10}^{-1}(f(z'), f(z), |z-z'|).\end{equation*}
Using the formula (ENS-4.6) from  \cite{kausel2012laplace} we obtain
\begin{equation}\begin{split}\label{eqn:L_elliptic}
\textbf{I}_{\rm 10}^{-1}(a,b,s)=&\frac{1}{\pi a}\left[\frac{2\sqrt{ab}}{\kappa}\textbf{E}(\kappa)+(a^2-b^2)\frac{\kappa}{2\sqrt{ab}}\textbf{K}(\kappa)
\right]\\
&+\frac{s}{\pi a} \textbf{Sgn}(a-b){\bf\Lambda}(\nu,\kappa)-\frac{s}{a}{\bf \Theta}(a-b),
\end{split}\end{equation}
where
\begin{equation}\begin{split}
\kappa=\frac{2\sqrt{ab}}{\sqrt{(a+b)^2+s^2}},\quad \nu=\frac{4ab}{(a+b)^2},\\ {\bf\Lambda}(\nu,\kappa)=\frac{|a-b|}{a+b}\frac{s}{\sqrt{(a+b)^2+s^2}}{\bf\Pi}(\nu,\kappa),
\end{split}\end{equation}
where
$\textbf{E}$ is the elliptic function of the first kind, $\textbf{K}$ is the elliptic function of the second kind, ${\bf\Pi}$ is the elliptic function of the third kind, $\textbf{Sgn}$ is the sign function,  and ${\bf\Theta}$ is the unit step function.

For the convenience of the reader, we recall
\begin{equation}
\label{eqn:elliptic}
\begin{split}
{\bf K}(\kappa)=&\int_{0}^{1}\frac{dt}{\sqrt{1-t^2}\sqrt{1-\kappa^2 t^2}},\\
{\bf E}(\kappa)=&\int_{0}^{1}\frac{\sqrt{1-\kappa^2t^2}}{\sqrt{1-t^2}}dt,\\
{\bf \Pi}(\nu,\kappa)=&\int_{0}^{1}\frac{dt}{(1-\nu t^2)\sqrt{1-t^2}\sqrt{1-\kappa^2 t^2}}.
\end{split}
\end{equation}

Substituting \eqref{eqn:letters} in \eqref{eqn:L_elliptic} we obtain the explicit expression
\begin{equation}\label{Y}
\begin{split}
\textbf{Y}(f(z'), f(z), \lvert z-z'\rvert)&= \frac{\sqrt{(z'-z)^2+(f(z')+f(z))^2}}{\pi f(z')}\\
&\cdot\textbf{E}\left(\frac{2\sqrt{f(z')f(z)}}{\sqrt{(z'-z)^2+(f(z')+f(z))^2}}\right)\\
& +\frac{f(z')^2-f(z)^2}{\pi f(z') \sqrt{(z'-z)^2+(f(z)+f(z'))^2}}\\
& \cdot \textbf{K}\left(\frac{2\sqrt{f(z') f(z)}}{\sqrt{(z'-z)^2+(f(z')+f(z))^2}}\right)\\
& +\frac{(z'-z)^2(f(z')-f(z))}{\pi f(z') (f(z')+f(z))\sqrt{(z'-z)^2+(f(z')+f(z))^2}}\\&\cdot {\bf\Pi} \left(\frac{4 f(z') f(z)}{(f(z')+f(z))^2},\frac{2 \sqrt{f(z') f(z)}}{\sqrt{(z'-z)^2+(f(z')+f(z))^2}}\right)\\
& -\frac{\lvert z'-z\rvert}{f(z')}{\bf\Theta}\left (f(z')-f(z)\right).
\end{split}
\end{equation}

Using equation \eqref{UGevaluates'2}, we obtain the following:
\begin{proposition}\label{prop:1} The gravitational potential at a point of cylindrical coordinates $(f(z),\phi,z)$ on the surface of a body generated by revolving the graph of $z\mapsto f(z)\geq 0$, $-z_0\leq z\leq z_1$ is given by
\begin{equation}\label{UGwY}
U_{\rm G}=-2 \pi G \rho \int_{-z_0}^{z_1} dz' \, f(z') \textbf{Y}(f(z'), f(z), \lvert z-z'\rvert).
\end{equation}
\end{proposition}

For a given body shape generated by the profile function $f(z)$, equation \eqref{UGwY} gives the gravitational potential $U_G$ as a function of $z$ at any point of the surface.

Equation \eqref{UGwY} can be easily modified to obtain the exact gravitational potential at any point in space, as follows:
\begin{corollary}\label{cor:1} The gravitational potential at a point in space of cylindrical coordinates $(s,\phi,z)$,  exerted by a body generated by  revolving the graph of $z\mapsto f(z)\geq 0$, $-z_0\leq z\leq z_1$,
%and symmetric with respect to the $s-$axis,
is given by
\begin{equation}\label{UGwYs}
U_{\rm G}=-2 \pi G \rho \int_{-z_0}^{z_1} dz' \, f(z') \textbf{Y}(f(z'), s, \lvert z-z'\rvert).
\end{equation}
\end{corollary}

The formulas  \eqref{UGwY} and \eqref{UGwYs} are very general as they apply to any solid of revolution.
They give the gravitational potential in terms of a $1$-dimensional integral of a combination elliptic functions.
It is known that the elliptic functions have expansions in power series that are convergent, thus \eqref{UGwY} and \eqref{UGwYs} can themselves be expanded in convergent power series \cite{byrd2013handbook}. Also, elliptic functions are readily implemented in many numerical computation software packages.

%Care must be taken when using the elliptic functions from standard programming languages such as C, Python, Matlab, Mathematica,  since the definition of the arguments does not follow a  uniform convention.
%For consistency, in the appendix we define the elliptic functions used in equation~\eqref{Y}.

\section{TOTAL ENERGY AT THE SURFACE OF A ROTATING AXISYMMETRIC
BODY }
\label{section2}
\begin{figure}[h]
\begin{center}
{\includegraphics[width=0.45\linewidth]{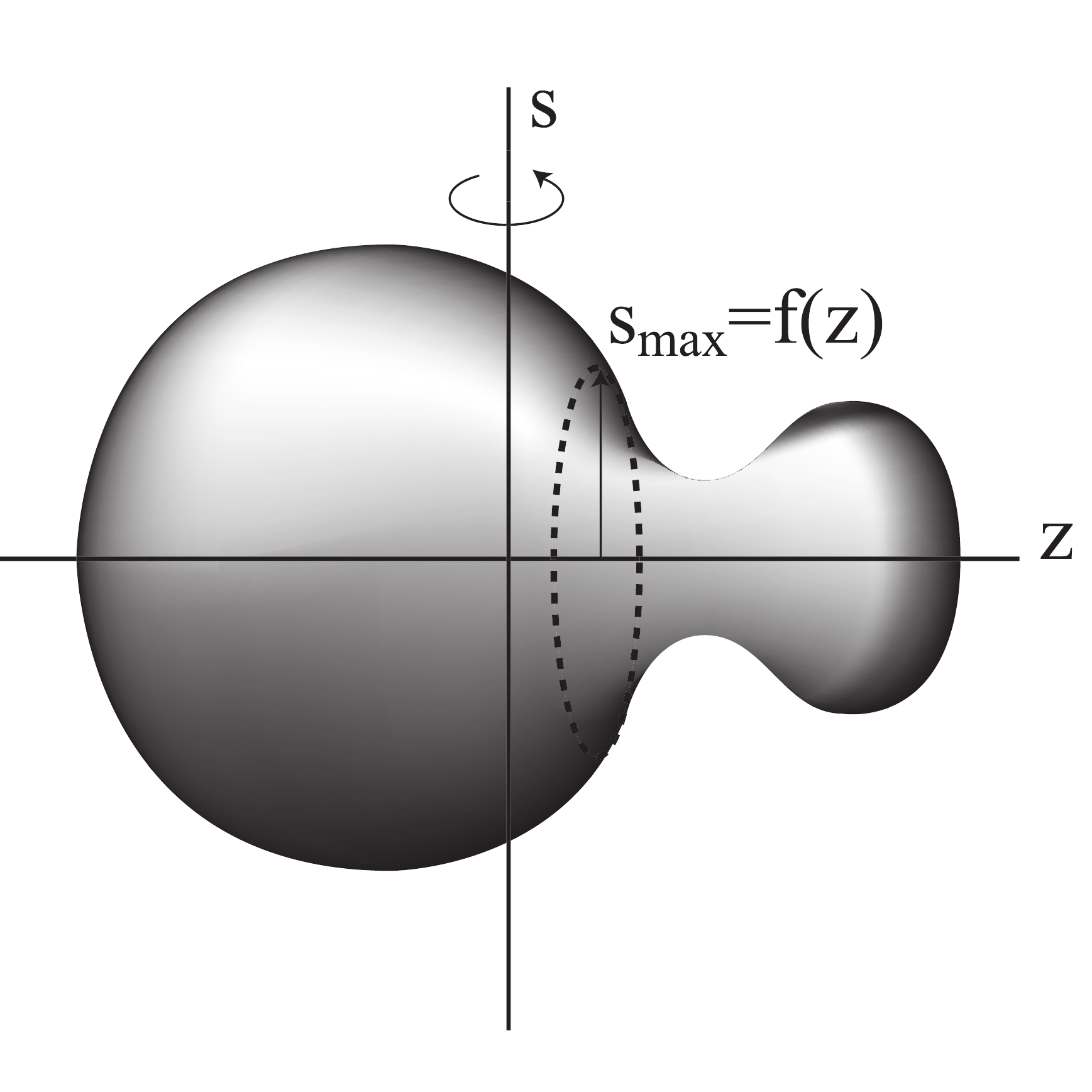}}
{\caption{Rotating dumbbell showing a circular region of surface points at $z$.}\label{fig:circularregionatz}}
\end{center}
\end{figure}

In this section we consider a solid of revolution as in  Section \ref{section1},
where the $s$-axis is chosen to pass through the center of mass.
We now  assume that the body rotates around the $s$-axis with constant  rotational speed $\omega$.
See Figure \ref{fig:circularregionatz}.
%The dumbbell rotates with constant angular speed $\omega$ around the $s$-axis.
In the reference frame of the rotating object, that is, one rigidly attached to the object, a centrifugal force ensues.  The magnitude of this force can be derived from the rotational kinetic energy.

To evaluate this rotational kinetic energy at any point on the surface we consider, as shown in Figure~\ref{fig:circularregionatz}, a ring of surface points at coordinate $z$.  A generic point on the ring has cylindrical coordinates $(f(z),\phi,z)$, so its distance to the $s$-axis is $f^2 (z) \sin^2 (\phi)+z^2$.   %gravitational
The rotational kinetic energy of a point of unit mass at the surface, which rotates around the $s$-axis with rotational speed $\omega$ is:
\begin{equation}\label{Urot}
U_{\rm Rot}=-\frac{1}{2} \omega^2 (f^2(z)\sin^2(\phi) +z^2).
\end{equation}
\begin{corollary}\label{corollary2}
The total energy $U$ acting on a particle of unit mass  at the surface of the rotating body  is the sum of the expressions in equations \eqref{UGwY}  and \eqref{Urot}:
\begin{equation}\label{U}
U=-2 \pi G \rho \int_{-z_0}^{z_1} dz'\, f(z') \textbf{Y}(f(z'), f(z), \lvert z-z'\rvert)-\frac{1}{2} \omega^2(f^2(z)\sin^2(\phi) +z^2).
\end{equation}
\end{corollary}

The negative sign in the last two terms of \eqref{U} accounts for the sign of the repulsive centrifugal force being opposite to that of the attractive gravitational force.

Equation \eqref{U} can be used in two ways.

First, if the explicit shape of a body is known, then the gravitational acceleration $g=-\nabla U$ at the surface of the body can be computed.

Second, one may ask, for a given family of shapes defined by a function $f(z)$ depending on parameters, to determine the parameters that (approximately) give the equilibrium  shape.  This can be re-phrased as a minimization problem, on determining the optimal parameters for the function describing the shape,  for which  the  total energy function at the surface has the lowest variability -- expressed as normalized standard deviation (see Section \ref{section3}).

Thus equation \eqref{U} can be seen as a main result of this paper.

In principle, for a given $\omega$, functional variations $\delta U$ should provide the function $f(z)$ that give the equilibrium shape.  However, in practice, that is a very challenging program, owing to the complicated form of the kernel of $\mathbf{Y}$.  Realistically one should explore the minimization problem in the parameter space of a well suited family of functions.  %In the next section we give an example of such a procedure applied to a parametrized family of dumbbell shapes.

A problem that immediately becomes apparent from equation \eqref{U} is that, at the surface of the object, $U_G$ depends only on  $z$, while $U_{Rot}$ depends, in addition, on $\phi$.  Thus, as one walks along the ring of figure \ref{fig:circularregionatz}, $U_G$ remains constant while $U_{Rot}$ contributes with an additional $\sin^2 \phi$ dependence.  It is clear that the total energy on a unit mass cannot be constant on the ring.  However, the practical problem of a real celestial body must be interpreted in the context of rotating not with respect to a fixed axes, but secularly with respect to all axes perpendicular to  the $z$-axis.  Under these conditions, and owing to the $\sin^2 \phi$  factor, outstretched shapes will develop perpendicularly to $z$.  But these shapes will subsequently develop in other directions, as the axis of rotation itself wobbles.  Hence, one should find among celestial bodies, those that after long times compared with the rotational period  $2\pi/\omega$, have cross sections averaged in $\phi$.
For this reason, we eliminate the dependence on $\phi$ by replacing $U_{\rm Rot}$ by its average with respect to $\phi$. Thus, instead of the total energy \eqref{U} we  consider the effective total energy

\begin{equation}\label{Ueff}
U_{\rm Eff}= \frac{1}{2\pi}\int_{0}^{2\pi} d\phi\, U(z, \phi),
\end{equation}
which takes the following form
\begin{equation}\label{Ueffexplicit}
U_{\rm Eff}=-2\pi G \rho \int_{-z_0}^{z_1} dz'\, f(z')  \textbf{Y}(f(z'), f(z), | z-z'|)-\frac{1}{4} f^2(z)\omega^2-\frac{1}{2}z^2 \omega^2.
\end{equation}

In the next section we give an example of
a parametrized family of dumbbell shapes, and investigate numerically the minimization problem
of finding parameters for which the effective total energy function
 \eqref{Ueffexplicit} at the surface has the lowest normalized standard deviation.

\section{MINIMIZATION PROBLEM FOR A PARAMETRIC FAMILY OF DUMBBELL SHAPES}
\label{section3}

\begin{figure}[H]
\begin{center}
  \begin{subfigure}[b]{0.465\textwidth}
    \includegraphics[width=\textwidth]{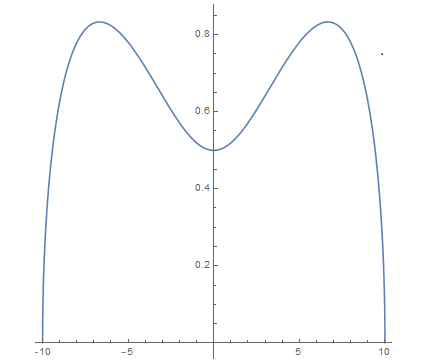}
    \caption{The function $f(z)$ at $z_0$ equals $10$, $\gamma$ equals $0.5$ and $\beta$ equals $0.9$. \vspace{4.2mm}}
    \label{fig:toyfunct1}
  \end{subfigure}
  \hspace{2mm}
  \begin{subfigure}[b]{0.495\textwidth}
    \includegraphics[width=\textwidth]{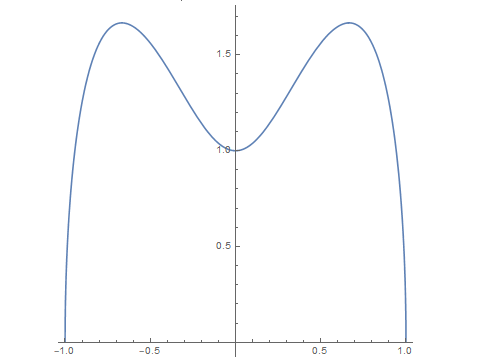}
    \caption{The function $f(z)$ at $z_0$ and $\gamma$ both equal $1$ (i.e. dimensionless) and $\beta$ equals $0.9$.}
    \label{fig:toyfunct2}
  \end{subfigure}
  \end{center}
  \caption{Examples of shapes given by the function $f(z)$, which is defined as in equation (\ref{toyeq}).}
\end{figure}

In this section we consider the family of curves
\begin{equation}\label{toyeq}
f(z)=\gamma \sqrt{\left(1-\left(\frac{z}{z_0}\right)^2\right)\left(1+\frac{\beta}{1-\beta}\left(\frac{z}{z_0}\right)^2\right)},
\end{equation}
for $-z_0\leq z\leq z_0$. The functions in this family are symmetric relative to the $s$-axis, so the solid of revolution generated by these functions have a mirror symmetry with respect to the plane through the origin orthogonal to the $z$-axis.
Different values of the parameters $\beta$ and $\gamma$ give different shapes.  The parameter $\gamma$ gives the value $f(0)$ of the radius of the section at the origin, and $\beta$ controls the convexity.
We note that some values of $\beta$ the resulting body is dumbell shaped while for some others it is not.
The parameter $z_0$ can be used as the unit of length and, without loss of generality, it can be set to unity. Examples of graphs of $f(z)$ obtained for some choices of parameters $(\beta, \gamma)$ are depicted in Figure~\ref{fig:toyfunct1} and Figure~\ref{fig:toyfunct2}.

%\begin{figure}[h!]
%\begin{center}
%{\includegraphics[width=0.35\linewidth]{Fig3a.png}}
%{\caption{The function $F(z)$ at $z_0$ equals $10$, $\gamma$ equals $0.5$ and $\beta$ equals $0.9$.}\label{fig:toyfunct1}}
%\end{center}
%\end{figure}
%\pagebreak
%\begin{figure}[ht!]
%\begin{center}
%{\includegraphics[width=0.35\linewidth]{Fig3b.png}}
%{\caption{The function $F(z)$ at $z_0$ and $\gamma$ both equal $1$(i.e. dimensionless) and $\beta$ equals $0.9$.}\label{fig:toyfunct2}}
%\end{center}
%\end{figure}

Replacing $f(z)$ from equation \eqref{toyeq} into equation \eqref{Ueffexplicit} and computing the integral numerically, we obtain numerical values of the function $U_{\rm Eff} (z,\beta,\gamma)$.  We then search, among all pairs $(\beta,\gamma)$, the ones  which produce, for a given $\omega$, an effective total energy   $U_{\rm Eff}$ with the least normalized standard deviation in $z$. The variant of the  normalized standard  deviation that we use is given by the  quantity $\sigma/|\mu|$, that is, the standard deviation divided by the mean. This quantity measures  the extent of variability in relation to the mean. It is known as the coefficient of variation, and it has been used extensively in the optimization literature  \cite{abdi2010coefficient}.
It is   a dimensionless quantity that is very practical in comparing the variability of different data sets. One particularity is that it emphasizes deviations for smaller means.

In our case, we compute  the coefficient of variation    $\sigma/|\mu|$ as a measure of  the variability of the effective total energy $U_{\rm Eff}$ computed for all values of $z$,  for different choices of parameters $(\beta,\gamma)$  which control the shape of the body. The optimization problem is to find those parameters for which the coefficient of variation attains the lowest values.

\begin{figure}[H]
\includegraphics[width=1\linewidth]{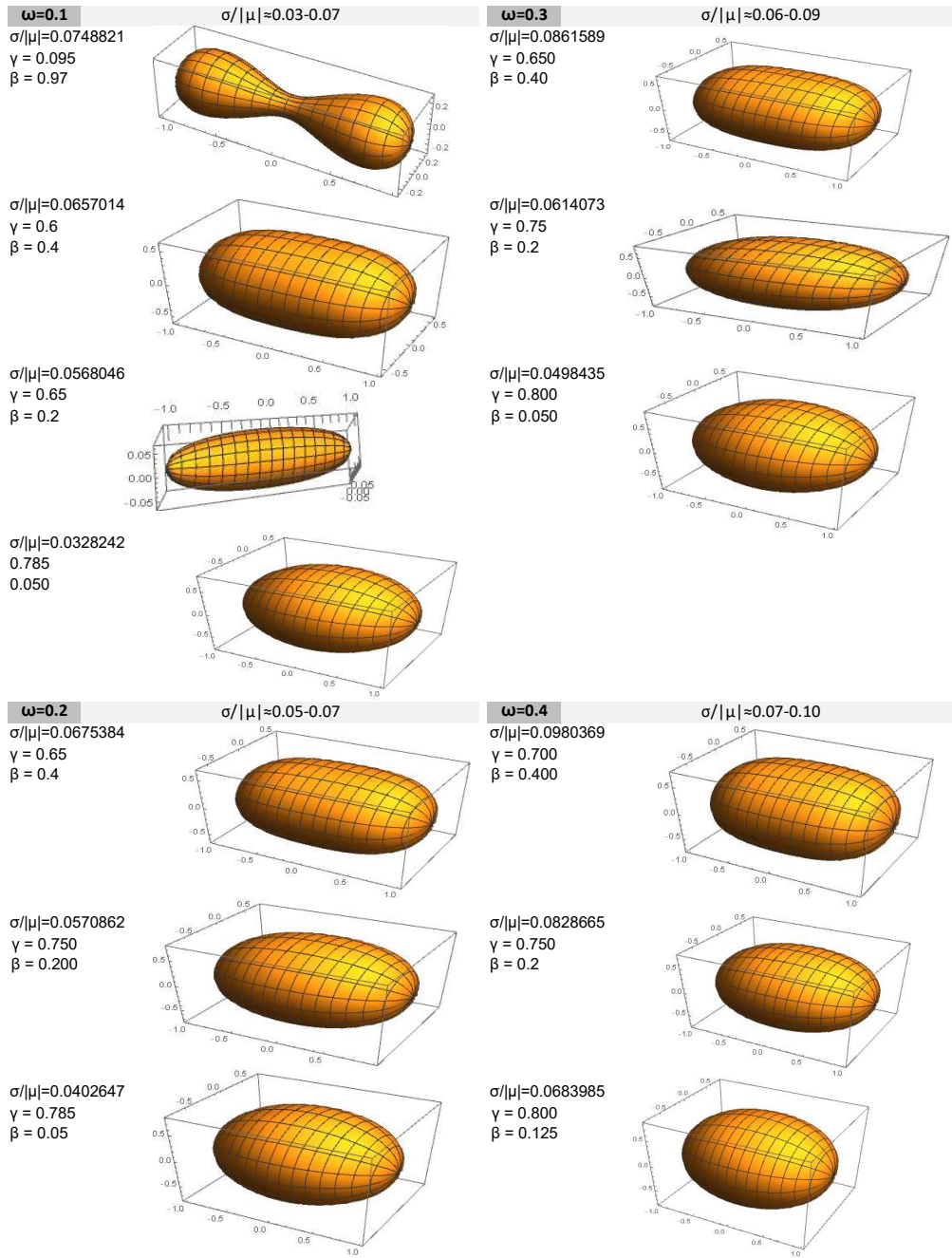}\caption{Approximate equilibrium shapes for $\omega=0.1$ and $\omega=0.2$, $\omega=0.3$ and $\omega=0.4$.}\label{fig:shapes1}
\end{figure}

\begin{figure}[H]
\includegraphics[width=1\linewidth]{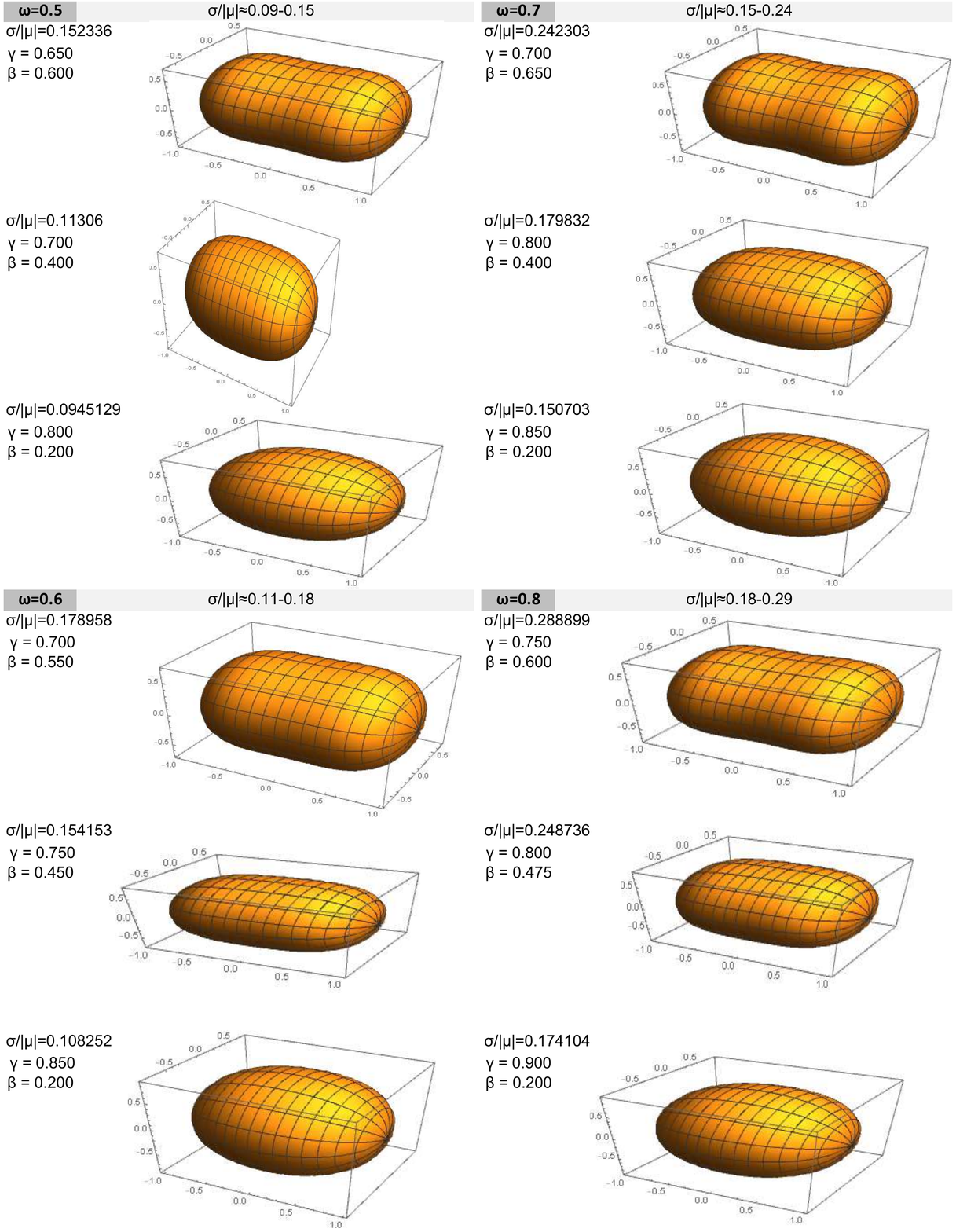}\caption{Approximate equilibrium shapes for $\omega=0.5$ and $\omega=0.6$, $\omega=0.7$ and $\omega=0.8$.}\label{fig:shapes2}
\end{figure}

\begin{figure}[H]
\newlength{\imagewidth}
\settowidth{\imagewidth}{\includegraphics{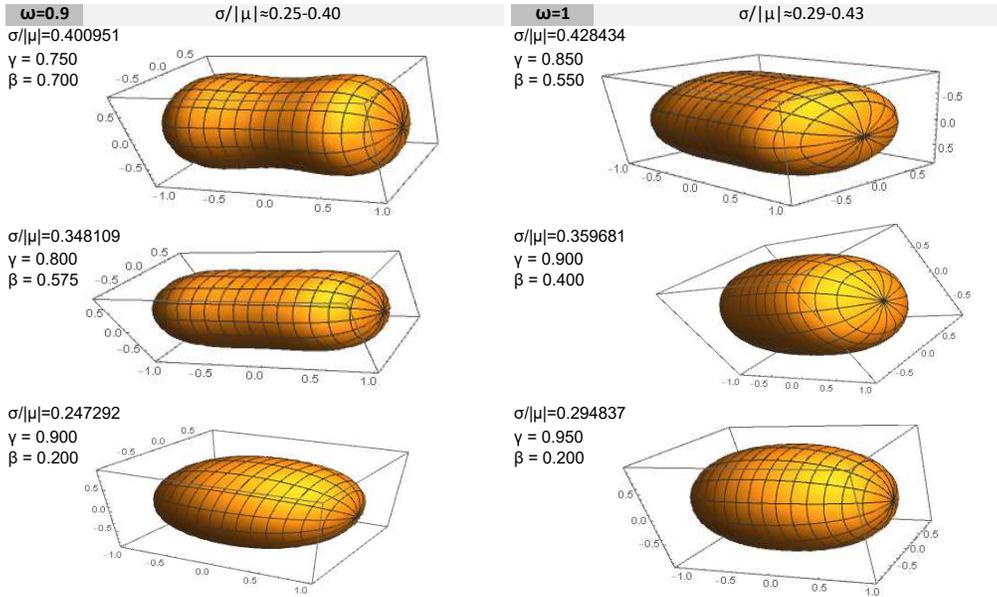}}
\includegraphics[trim=0 0.65\imagewidth{} 0 0, clip, width=1\linewidth]{omega091.pdf}\caption{Approximate equilibrium shapes for $\omega=0.9$ and $\omega=1.0$.}\label{fig:shapes3}
\end{figure}

Practically,  for each value of the rotational speed $\omega$ in a grid, we compute the effective total energy  for  each coordinate $z$, for  fixed $(\beta,\gamma)$. The goal is to find  dumbbell shapes that yield  a nearly  constant effective total energy at the surface, in practice one with a relatively small coefficient of variation of $U_{\rm Eff}$ over the surface of the body. For each value of $\omega$ -- which we increase at each step by an increment of $\delta \omega=0.1$,  we record the local minimum values of $\sigma/|\mu|$ and the parameters  $\gamma$ and $\beta$ for which the local minimum is attained.
The obtained results suggest different interesting dumbbell shapes, as in Figures~\ref{fig:shapes1},
~\ref{fig:shapes2},~\ref{fig:shapes3}.

%\begin{figure}
%\begin{tabular}{p{0.5\textwidth} p{0.5\textwidth}}
%\vspace{0pt}\includegraphics[width=0.89\linewidth]{omrgaone.png}
%&\vspace{0pt}\includegraphics[width=0.90\linewidth]{omega2.png}\\
%\vspace{0pt}\includegraphics[width=0.90\linewidth]{omega3.png}&
%\vspace{0pt}\includegraphics[width=0.90\linewidth]{omega4.png}
%\end{tabular}
%\caption{Approximate equilibrium shapes for $\omega=0.1$ and $\omega=0.2$, $\omega=0.3$ and $\omega=0.4$.}
%\label{fig:shapes1}
%\end{figure}

%\begin{figure}
%\begin{tabular}{p{0.5\textwidth} p{0.5\textwidth}}
%\vspace{0pt}\includegraphics[width=0.9\linewidth]{omega5.png}&
%\vspace{0pt}\includegraphics[width=0.9\linewidth]{omega6.png}\\
%\vspace{0pt}\includegraphics[width=0.9\linewidth]{omega7.png}&
%\vspace{0pt}\includegraphics[width=0.9\linewidth]{omega8.png}\\
%\end{tabular}
%\caption{Approximate equilibrium shapes for $\omega=0.5$ and $\omega=0.6$, $\omega=0.7$ and $\omega=0.8$.}
%\label{fig:shapes2}
%\end{figure}

%\begin{figure}
%\begin{tabular}{p{0.5\textwidth} p{0.5\textwidth}}
%\vspace{0pt}\includegraphics[width=0.9\linewidth]{omega9.png}&
%\vspace{0pt}\includegraphics[width=0.9\linewidth]{omega11.png}
%\end{tabular}
%\caption{Approximate equilibrium shapes for $\omega=0.9$ and $\omega=1.0$.}
%\label{fig:shapes3}
%\end{figure}

We notice that, for the same $\omega$, there may be several different choices of parameters $\gamma$ and $\beta$ -- hence different shapes -- for which $\sigma/|\mu|$ attains a local minimum. Since at a  practical level we do not
look for exact solutions of the optimization problem, but for approximate ones for which $\sigma/|\mu|$ is `relatively small', we do not impose a precise threshold on what `relatively small' means. However, we speculate that for a given $\omega$,  dumbbell shapes with relatively large  values of $\sigma/|\mu|$  are less likely to occur in real life than those with relatively small values   of $\sigma/|\mu|$.
We also notice that some of the shapes that we obtain appear to be similar to the observed shapes of some  asteroids and comets, such as 624 Hektor, 103P/Hartley, and  8P/Tuttle.
% Wai-Ting:  please do a best fit of these real comets to our parametric families to see what parameters we get

\section{CONCLUSIONS}
\label{section4}
First,  we have derived a formula in terms of elliptic integrals for the gravitational potential at any point  on the surface of an axisymmetric body, as well as at any point outside the body.
Second, we have derived a formula for the total energy  of an unit mass particle on the surface of an axisymmetric
body that rotates around an axis perpendicular to the symmetry axis.
Third, we have formulated an optimization problem of finding approximate equilibrium shapes, based on the principle of minimizing the coefficient of variation  of the effective total energy at the surface.
As an application, we have considered a two-parameter family of dumbbells, and computed numerically their  approximate equilibrium solutions,   i.e., the choices of parameters, depending on the rotational speed, for which the effective total energy at the surface is approximately constant.

We note that there also exist exact equilibrium solutions of dumbbell shapes \cite{eriguchi1982dumb}. Such dumbbell shapes are not given by closed form equations. In contrast, we provide a family of dumbbell shapes that are given by simple, explicit formulas, and depend only on two parameters. However these only correspond to approximate energy level sets. Our family of dumbbell shapes could be potentially utilized to find first approximations for irregularly shaped asteroids and comets. Our approach can be extended to other families of shapes (depending on more parameters), as well as to shapes that are not generated as solids of revolution.
%Moreover, we can derive formulas for the gravitational potential generated by such shapes at any point in space.

\bibliographystyle{apacite}
\bibliography{Ref}

\end{document}